\documentclass[runningheads]{llncs}

\usepackage[T1]{fontenc}
\usepackage{graphicx} 
\usepackage[backend=biber, style=phys, sorting=none]{biblatex}
\usepackage{hyperref}
\usepackage{amsmath}
\usepackage{amsfonts}
\usepackage{booktabs}
\usepackage{makecell}
\usepackage{subfig}
\usepackage{bbold}
\usepackage{amssymb}
\usepackage[margin=1.2in]{geometry}
\usepackage{textcomp}
\usepackage{physics}
\hypersetup{colorlinks,%
  citecolor=blue,%
  filecolor=black,%
  linkcolor=blue,%
  urlcolor=blue}

\title{Entanglement and discord classification via deep learning}
\titlerunning{Entanglement and discord classification via deep learning}

\author{Katherine Mu\~noz-Mellado\inst{1}\orcidID{0009-0006-2628-0321} \and Daniel Uzc\'ategui-Contreras\inst{2}\orcidID{0009-0000-0375-6549} \and Antonio Guerra\inst{1}\orcidID{0009-0007-3624-9755} \and Aldo Delgado\inst{1}\orcidID{0000-0002-8968-5733}\and Dardo Goyeneche\inst{3}\orcidID{0000-0002-9865-4226}}

\authorrunning{D. Uzc\'ategui-Contreras, D. Goyeneche et al.}

\institute{
\textit{Departamento de F\'isica e Instituto Milenio de Investigaci\'on en \'Optica},
Universidad de Concepci\'on, Casilla 160-C, Concepci\'on, Chile\\
\email{katherinepmunoz@udec.cl, antongonzalez@udec.cl, aldelgado@udec.cl}\and
\textit{Departamento de Matem\'atica y F\'isica Aplicadas},
Universidad Cat\'olica de la Sant\'isima Concepci\'on, Alonso de Ribera 2850, Concepci\'on, Chile\\
\email{daniel.uzcategui@ucsc.cl}\and
\textit{Instituto de F\'isica}, 
Pontificia Universidad Cat\'olica de Chile, Casilla 306, Santiago, Chile\\
\email{dardo.goyeneche@uc.cl}
}

\begin{document}

\maketitle      

\pagestyle{headings}

\begin{abstract}

In this work, we propose a deep learning-based approach for quantum entanglement and discord classification using convolutional autoencoders. We train models to distinguish entangled from separable bipartite states for $d \times d$ systems with local dimension $d$ ranging from two to seven, which enables identification of bound and free entanglement. Through extensive numerical simulations across various quantum state families, we demonstrate that our model achieves high classification accuracy. Furthermore, we leverage the learned representations to generate samples of bound entangled states—the rarest form of entanglement and notoriously difficult to construct analytically. We separately train the same convolutional autoencoders architecture for detecting the presence of quantum discord and show that the model also exhibits high accuracy while requiring significantly less training time.
\end{abstract}

\section{Introduction}

Quantum resource theories (QRTs) provide a framework for the quantitative analysis of quantum resources \cite{PhysRevLett.115.070503,PhysRevLett.118.060502,RevModPhys.91.025001}. Quantum phenomena such as entanglement \cite{PhysRevLett.90.100402}, coherence \cite{RevModPhys.89.041003}, purity \cite{PhysRevA.67.062104}, symmetry (or lack thereof) \cite{Marvian2014} and thermodynamics \cite{PhysRevLett.111.250404} have been studied from the point of view of QRT, providing an accurate picture of the role played by these resources in a wide range of quantum tasks involving device-independent distribution of a secret key, quantum communications, and quantum state discrimination \cite{PhysRevResearch.6.043303,e23010073,e21030263}, among others. 

One of the most intensively and deeply studied quantum resources is quantum entanglement, a phenomenon highlighted by Einstein, Podolsky, and Rosen \cite{epr_1935} and formally introduced by Schr\"odinger \cite{schrodinger_1935}, which is widely recognized as one of the most distinctive features of quantum mechanics. Beyond its foundational relevance, entanglement plays a central role as a resource in numerous quantum processes and applications, including quantum cryptography \cite{ekert_1991}, quantum teleportation \cite{bennet_1993}, quantum communication \cite{bennet_1992}, and quantum metrology \cite{huang_2024}.

A central challenge in the study of entanglement is determining whether a given density matrix describes a separable or entangled state. This problem becomes especially difficult for mixed quantum states, for which efficient separability criteria are generally hard to obtain \cite{gurvits_2003, gharibian_2010}. Among the most prominent tools is the positive partial transpose (PPT) criterion, also known as the Peres-Horodecki criterion \cite{peres_1996, horodecki_1996}, which partially characterizes entanglement in bipartite systems. The bipartite states with negative partial transpose (NPT) are entangled. However, the converse is not generally true since there are entangled PPT states. In addition to its effectiveness in detecting entanglement, the PPT criterion has proven useful in quantifying entanglement \cite{vidal_2002} and identifying distillable states \cite{horodecki_1998_bound_entanglement}. However, it falls short of providing a definitive answer to the separability problem, as it cannot distinguish separable states from bound entangled states, that is, entangled states that are not distillable. Other widely used methods for entanglement detection include entanglement witnesses \cite{lewestein_2000} and the computable cross-norm or realignment criterion (CCNR) \cite{lewestein_2000, chen_2003}, among others.

In addition to quantum entanglement, there are other features of quantum mechanics that can play the role of a resource. Quantum discord is defined as the measure of quantum correlation in a bipartite system that quantifies the degree to which the total correlation of the system deviates from classical correlations \cite{ollivier2001}. For pure states, quantum discord is equal to entanglement. However, a bipartite quantum system in a mixed state may not have entanglement and yet exhibit quantum discord. A related quantity is quantum dissonance, which is the portion of quantum discord that remains even after removing entanglement. One process based on quantum dissonance is assisted optimal state discrimination (AOSD) \cite{Zhang2013}, where two non-orthogonal quantum states are discriminated by a three-outcome positive operator-valued measure (POVM). This is usually implemented by introducing an auxiliary system (ancilla), coupling it to the main system, and performing a joint projective measurement. As with entanglement, quantum discord is notoriously difficult to calculate, as it involves an optimization procedure on the set of projective measurements. Indeed, determining quantum discord falls under NP-completeness \cite{Huang2014}, meaning that, for most quantum systems, assessing quantum discord is viewed as computationally challenging because the complexity of the algorithms required is thought to increase exponentially with the system's dimension. Examples of analytical results for the value of quantum discord are only known for certain classes of quantum states with high symmetry \cite{Rau2017}.\\

In this work, we employ a deep learning-based approach to study the problem of whether a given bipartite quantum state exhibits entanglement or discord. In particular, we consider the case of a two-qudit system, where the dimension of both qudits is equal to $d \in \{2, \dots, 7\}$. First, we focus on entanglement. We present an unsupervised deep learning model, based on convolutional autoencoders (CAE), for entanglement classification. This model is trained exclusively on separable states that are randomly generated as convex combinations of tensor products of mixed local states. In particular, the reconstruction error induces a clear separation between entangled NPT states and mixed separable states. Interestingly, the model also naturally infers the invariance of entanglement under local unitary transformations, a feature that is not explicitly included in the loss function. Therefore, the classifier's predictions remain invariant under random local unitary transformations applied to either mixed separable states or NPT states. We also tested the model on bound entangled states (entangled PPT states) for $d \in \{3, \dots, 7\}$. Initially, these states were incorrectly classified as separable, which we traced to a distribution mismatch: the training data consisted of quantum states with dense matrix representations, while the bound entangled states we tested had sparse representations. To address this, we exploited the invariance of entanglement under local unitary transformations, converting sparse matrices to dense ones before classification. This preprocessing led to correct classification of bound entangled states with high success rates across all dimensions tested. Building on this classifier, we developed a procedure to generate new bound entangled states for $d \in \{3, \dots, 7\}$, which we certified using the symmetric extension entanglement criterion \cite{doherty_2004}. Our classifier achieves near-perfect performance for dimensions $d \geq 3$, with accuracy consistently exceeding $98\%$ across all tested ensembles (including NPT, mixed separable state and entangled states with PPT).

We also trained a model to identify states with quantum discord. This model is trained using only classical-classical states, which are separable but without discord. These states are convex combinations of tensor products of pure local states. This model allows for a highly accurate classification of the quantum-classical and classical-quantum states, which are separable but exhibit quantum discord. Interestingly, this model, unlike the entanglement classifier, requires a single epoch to achieve classification accuracies above $99\%$ for $d \geq 3$.

\subsection*{Related work}
Previous studies have demonstrated the potential of machine learning techniques in detecting entanglement for multiqubit systems \cite{lu_2018, Chen_2022, huang_2025, asif_2023, urena_2024}, with some achieving accuracies exceeding $97.5\%$ for systems with up to 10 qubits. The case of two qutrits was also investigated \cite{lu_2018}, combining machine learning with convex hull approximations. Furthermore, machine learning techniques have been employed to estimate the volume of various types of entanglement, including bound entanglement ~\cite{Hiesmayr_2021}.

Other studies have used autoencoders and convolutional neural networks (CNNs) for entanglement classification. For example, a complex-valued network architecture has been developed that combines pseudo-siamese blocks and generative adversary networks (GANs) \cite{Chen_2022}. Although this model is trained on multiqubit separable mixed states, the architecture is structurally complex; in contrast, our approach is significantly more streamlined, utilizing a single-stage CAE. We employ the reconstruction error—measured by a matrix norm—as a direct unsupervised indicator of entanglement for bipartite systems. Furthermore, unlike previous detection-focused studies, we leverage this architecture for the active generation of novel bound entangled states. Additionally, a supervised approach using a $\beta$-Variational Autoencoder \cite{higgins2017beta} has been proposed to classify the entanglement of two-qubit states based on the results of Pauli measurements \cite{nahum_2021}. Our method differs in that it is strictly unsupervised and operates directly on the density matrix, allowing us to capture the fundamental mathematical structure of separability and scale the analysis to higher-dimensional $d \times d$ systems ($d \leq 7$)

Regarding quantum discord, CNNs were used to detect quantum discord from simulated measurement data in two-qubit systems \cite{Taghadomi2025}. In this model, the architecture is trained to learn the optimal observables (encoded in the CNN's kernels), which enables a fully connected layer to perform the detection of discord. One version of this model, termed Branching CNN (BCNN), uses a restricted set of convolutional paths rather than the entire set, which implies a reduction in the number of resources required for discord detection

\section{Review on entanglement}

In this section, we briefly review the mathematical characterization of quantum states and the notions of separable and entangled states. We also review entanglement distillation, that is, the possibility of generating a maximally entangled state from a set of less entangled states by local operations and classical communication, and its relation the the phenomenon of bound entanglement. Finally, we state the partial transposition criterion for detecting entangled states and the limitations of this criterion.

\subsection{Quantum states}
The physical state of a quantum system is completely described by the density operator $\rho\in\mathcal{L}(\mathcal{H})$, which is a normalized ($\mathrm{Tr}(\rho)=1$) positive semi-definite ($\rho \ge 0$) linear operator acting on a Hilbert space $\mathcal{H}$. A quantum state is called \textit{pure} if $\mathrm{Tr}(\rho^2)=1$ and \textit{mixed} if $\mathrm{Tr}(\rho^2)<1$. For a system formed by $N$ bodies, each with local Hilbert space $\mathcal{H}_i$ of dimension $d_i$, the global Hilbert space is denoted $\mathcal{H} = \mathcal{H}_1 \otimes \ldots \otimes \mathcal{H}_N$. The density operator has a matrix representation, called the \textit{density matrix}.

\subsection{Entanglement}

Entanglement is a quantum correlation between parts of a system that cannot be explained by classical physics. A pure bipartite state $|\psi\rangle \in \mathcal{H}_A \otimes \mathcal{H}_B$ is entangled if it cannot be written as a product state {$|\phi_A\rangle \otimes |\phi_B\rangle$}, whereas a mixed state $\rho \in \mathcal{L}(\mathcal{H}_1 \otimes\ldots\otimes\mathcal{H}_N )$ is called separable if it can be written as a convex combination of product states

\begin{equation}
\label{eq:separable_state}
\rho = \sum_{i=1}^M p_i \rho_{i}^{1} \otimes \ldots \otimes \rho_{i}^{N}, \text{  where   } p_i \geq 0 \text{  and  } \sum_i p_i = 1.
\end{equation}

For bipartite systems ($N=2$) Eq.~\eqref{eq:separable_state} becomes $\rho = \sum_i p_i \rho^A_i \otimes \rho^B_i$. If a quantum state does not admit an expansion in the form of Eq.~\eqref{eq:separable_state}, then the state is \textit{entangled}. Detecting whether a general state is entangled or separable is a fundamental problem in quantum information theory, especially because many quantum protocols rely on the presence of entanglement \cite{ekert_1991, bennet_1993, bennet_1992, huang_2024}. Several criteria have been developed for this task, yet most of them are computationally expensive or are only valid for low-dimensional systems \cite{peres_1996, horodecki_1996, lewestein_2000, chen_2003}. This motivates the development of alternative strategies, such as those based on machine learning \cite{lu_2018, Chen_2022, huang_2025, asif_2023, urena_2024, Hiesmayr_2021, higgins2017beta, nahum_2021, Sabiote2026}.

\subsection{Entanglement distillation}

Let us consider a situation in which $n$ copies of a bipartite quantum state $\rho_{AB}$ are shared between two distant parties, Alice and Bob; one part of each copy is in the possession of Alice and the other in the possession of Bob, and all copies are distributed identically. \textit{Entanglement distillation} is the process of transforming these $n$ copies of $\rho_{AB}$ into $m$ copies of the maximally entangled state $|\Phi^{-}\rangle \langle \Phi^{-}|$ by means of \textit{local operations and classical communication (LOCC)} \cite{bennet_purification_1996}. This is usually denoted as
\begin{equation}
    \underbrace{\rho_{AB}^{\otimes n}}_{n \text{ copies}} \xrightarrow{\text{LOCC}} |\Phi^{-}\rangle \langle \Phi^{-}|^{\otimes m},
\end{equation}
where $|\Phi^{-}\rangle = \frac{1}{\sqrt{2}}\left(|0\rangle\otimes|1\rangle - |1\rangle\otimes|0\rangle\right)$ is a maximally entangled two-qubit Bell state. Since LOCC cannot transform separable quantum states into entangled ones, the only states suitable for distillation are entangled states \cite{horodecki_1999}. Given that many tasks in quantum information theory use maximally entangled states, distilling mixed states is relevant in situations where pure states such as $|\Phi^{-}\rangle$ are difficult to generate or protect against decoherence. 

However, it has been shown that not all mixed quantum states $\rho$ can be distilled into $|\Phi^{-}\rangle$ \cite{horodecki_1998_bound_entanglement}. Due to this, we now distinguish between two types of entanglement: \textit{free} entanglement, associated with mixed states that can be distilled into $|\Phi^{-}\rangle$, and \textit{bound} entanglement, associated with mixed states that cannot be distilled into a maximally entangled state.

\subsection{Partial positive transpose criterion}

In general, a bipartite state $\rho_{AB} \in \mathcal{L}(\mathcal{H}_A \otimes\mathcal{H}_B )$ is separable if and only if $(\mathcal{E}\otimes\mathbb{1}_{B})(\rho_{AB}) \geq 0$ for all positive maps $\mathcal{E}:\mathcal{L}(\mathcal{H}_A) \rightarrow \mathcal{L}(\mathcal{H}_A) $. An example of such a positive map is the transposition operation, denoted $T$. The \textit{partial transposition} of $\rho_{AB}$ with respect to the subsystem $A$ is defined as $\rho_{AB}^{T_A}= (T \otimes \mathbb{1}_{B})(\rho_{AB})$. If $\rho_{AB}^{T_A} \geq 0$ all its eigenvalues are non-negative, then $\rho_{AB}$ is said to be \textit{positive under partial transposition} (PPT). If $\rho_{AB}^{T_A}$ has at least one negative eigenvalue, then $\rho_{AB}$ is \textit{negative under partial transposition} (NPT) and implies that $\rho_{AB}$ is entangled. This is called the PPT criterion and is perhaps the simplest criterion to determine whether a mixed bipartite quantum state is entangled \cite{peres_1996,horodecki_1996}.

If $\rho_{AB}$ is the state of a two-qubit or a qubit-qutrit system, then the PPT criterion is necessary and sufficient for separability, that is, $\rho_{AB}^{T_A} \geq 0$ if and only if $\rho_{AB}$ is separable. However, in higher dimensions, there are entangled states such that $\rho^{T_A} \geq 0$. Furthermore, it has been shown that PPT states are undistillable \cite{horodecki_1996} and therefore entangled PPT states are so-called \textit{bound entangled states}. Whether there also exist undistillable states with NPT is an open question \cite{horodecki2022five}.

\section{Review on quantum discord}

In this section, we review the definition of quantum discord and show some families of bipartite separable quantum states with and without quantum discord, such as the classically correlated states that will be used to train the discord classifier. 

\subsection{Quantum discord}

Traditional entanglement metrics capture the totality of quantum correlations for pure states. However, for mixed states, the characterization of quantum correlations requires a more general approach, as there exist separable (non-entangled) states that still manifest non-classical quantum behavior and serve as essential resources for certain quantum information processing tasks. In this context, quantum discord ($Q$) emerges as a different measure of quantum correlation.

The concept of quantum discord \cite{ollivier2001,henderson2001} is formally defined as the difference between total correlations, quantified by quantum mutual information ($\mathcal{I}$), and classical correlations ($\mathcal{J}$), which represent the maximum classical information extractable through local measurements on a subsystem. From the perspective of the subsystem $A$ quantum discord is expressed as
\begin{equation}
     Q^A(\rho^{AB}) = \mathcal{I}(\rho^{AB}) - \mathcal{J}^A(\rho^{AB}),
\end{equation}
where $\mathcal{I}(\rho^{AB})$ represents the total quantum mutual information of the bipartite state $\rho^{AB}$, given by
\begin{equation}
    \mathcal{I}(\rho^{AB}) = S(\rho^A)+S(\rho^B)-S(\rho^{AB}),
\end{equation}
with $S(\rho)$ denoting the von Neumann entropy
\begin{equation}
    S(\rho)= -\mathrm{Tr}(\rho\log\rho).
\end{equation}
The classical correlation term $\mathcal{J}^A(\rho^{AB})$ is obtained by maximizing over all possible local projective measurements $\Pi^A$ in the subsystem $A$, that is,
\begin{equation}
    \mathcal{J}^A(\rho^{AB}) = \max_{\Pi^A} \mathcal{J}_{\Pi^A}(\rho^{AB}),
\end{equation}
where the classical correlation for a given measurement basis is defined as
\begin{equation}
    \mathcal{J}_{\Pi^A}(\rho^{AB}) = S(\rho^B) - \sum_i p_i^A S(\rho^{B|A=i}_{\Pi^A}).
\end{equation}
Here, $p_i^A$ represents the probability of obtaining the result $i$ when measuring the subsystem $A$, and $\rho^{B|A=i}_{\Pi^A}$ denotes the post-measurement state of the subsystem $B$ conditioned on the result of measurement $i$ in the subsystem $A$.

\subsection{Separable states with quantum discord}

Analogously to Eq.~(\ref{eq:separable_state}), which describes separable states $\sigma_{sep}^{AB}$ as convex combinations of tensor products of local states, it is possible to characterize the set of states lacking quantum discord. This set is formed by all states of the form
\begin{equation}
        \chi_{cc}^{AB} = \sum_{i=1}^{d_A} \sum_{j=1}^{d_B} p_{ij}^{AB} \ket{i^A}\bra{i^A}\otimes\ket{j^B}\bra{j^B}.
        \label{eq:cc}
\end{equation}
These are called classical-classical states (CC) and are within the set of separable states. The set $\{p_{ij}^{AB}\}$ is a joint probability distribution and $\{\ket{i^A}_{i=1}^{d_A}\}$ and $\{\ket{j^B}_{j=1}^{d_B}\}$ are orthonormal bases of $\mathcal{H}_A$ and $\mathcal{H}_B$, respectively. CC states are considered a stereotype of classicality within the quantum world. A key property is that the marginal states derived from a CC state remain classical states, that is, mixtures of orthonormal basis elements. Furthermore, $\chi_{cc}^{AB}$ states are exactly those that can be perfectly broadcast locally.

It is possible to define two sets of separable states with non-vanishing quantum discord. The classical-quantum states (CQ) $\chi_{cq}^{AB}$ are defined as
\begin{equation}
            \chi_{cq}^{AB} = \sum_{i=1}^{d_A} p_{i}^{A} \ket{i^A}\bra{i^A} \otimes \rho_i^{B}.
            \label{eq:cq}
\end{equation}     
This formulation embeds a classical state for subsystem A into the bipartite quantum state space. The set $\{p_i^A\}$ is a probability distribution, and $\{\ket{i^A}_{i=1}^{d_A}\}$ is an orthonormal basis of $\mathcal{H}_A$. The states $\rho_i^B$ are arbitrary states of subsystem B. In a CQ state, the marginal state of A ($\chi^A_{cq}$) remains classical. Critically, a bipartite state $\rho^{AB}$ is CQ if and only if the quantum discord, quantified by measurement in subsystem A ($D^{\to}(\rho^{AB})$ or $Q_I^A(\rho^{AB})$), is zero. Operationally, the CQ states are exactly the states that allow unidirectional local broadcasting (from A). They can also be termed incoherent-quantum states on some local basis on A.

Similarly, it is possible to define quantum-classical states (QC) $\chi_{qc}^{AB}$. This is the counterpart to the CQ set of states, where the classical component resides in subsystem B, and is defined as
\begin{equation}
        \chi_{qc}^{AB} = \sum_{j=1}^{d_B} p_{j}^{B} \rho^{A}_j \otimes \ket{j^B}\bra{j^B},  
        \label{eq:qc}
\end{equation}   
where $\{p_j^B\}$ is a probability distribution, $\{|\mathrm{j}^B\rangle\}$ is an orthonormal basis of $\mathcal{H}_B$, and $\rho_j^A$ are arbitrary states of subsystem A. A bipartite state $\rho^{AB}$ is quantum-classical if and only if the quantum discord, quantified by measurement on subsystem B, $Q_I^B(\rho^{AB})$ is zero.
        
The states belonging to the classical-classical, classical-quantum, and quantum-classical sets are crucial for exploring quantum correlations beyond entanglement. These classes of states are part of the definition of the quantum-classical boundary of correlations. Collectively, they are often referred to as classically correlated states. It is noteworthy that these three sets of states (CC, CQ, QC) form non-convex sets and occupy a region of zero measure (measure zero) in the space of all bipartite quantum states.

\section{Model}

In this section, we present an unsupervised deep-learning model for the classification of entanglement. Our method is based on CAEs trained exclusively on separable bipartite density matrices. For an input $\mathbf{x}$, an autoencoder is a deep learning map formed by two functions, an encoder, denoted $E(\ldots)$, and a decoder, denoted $D(\ldots)$. The autoencoders aim to reconstruct the input, that is, $\mathbf{x} \approx D_{\theta}(E(\mathbf{x}))$, where $\theta$ are the weights of the network.

CAEs are a particular type of autoencoder \cite{goodfellow-et-al-2016} consisting mainly of CNNs \cite{oshea2015introductionconvolutionalneuralnetworks}, pooling operations and activation functions. CNN represent a highly effective class of neural networks, commonly applied to multi-dimensional array data such as images. CNN excel at computer vision applications, including object detection, image classification, and segmentation tasks. CNN have also been implemented to study problems in quantum mechanics \cite{liang_2018,liang_2021,Miles2021}, including quantum entanglement in multi-qubit systems \cite{Qu2023}  and quantum discord in two-qubit systems \cite{Taghadomi2025}.  

\begin{figure}
    \centering
    \includegraphics[scale=0.37]{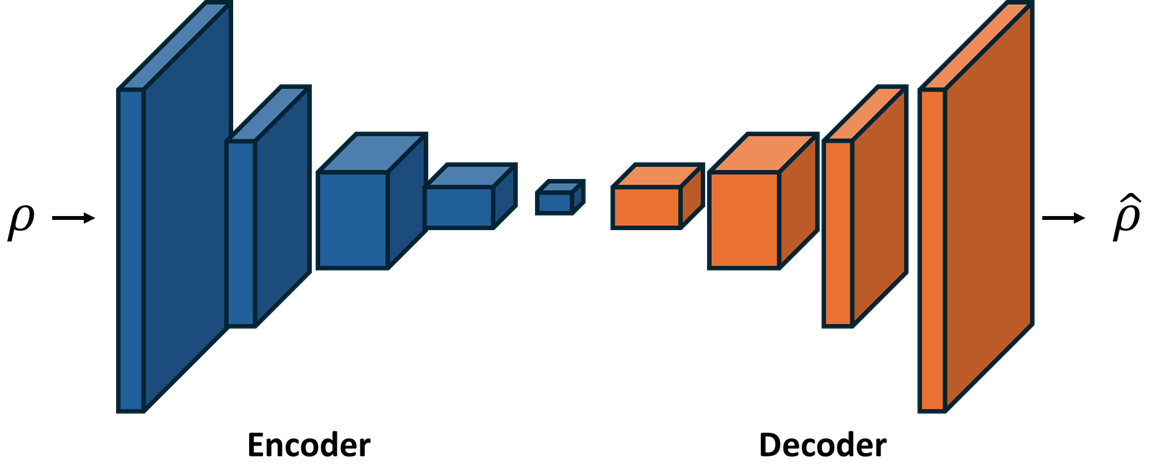}
    \caption{Schematic of the CAE architecture. The input density matrix $\rho$ undergoes dimensionality reduction through the encoder $E$ to a latent representation. The decoder $D$ then maps this representation back to the original Hilbert space dimensions to produce the reconstructed state $\hat{\rho} = D_\theta(E(\rho))$, where $\theta$ represents the weights of the entire network.}
    \label{fig:autoencoder}
\end{figure}

Figure \ref{fig:autoencoder} illustrates CAE's architecture. The encoder compresses the input through successive layers into a lower-dimensional representation, which the decoder then uses to reconstruct the input's original dimensions. The details of the architecture we implemented are given in appendix 1. We encode quantum states as two-channel multi-arrays, where one channel corresponds to the real part of the state and the other to the imaginary part. 

\subsection*{Loss function}

To train our CAEs we use the following loss function
\begin{equation}
    \label{eq:norm}
    L(\rho, \hat{\rho}) = \dfrac{1}{N_S}\sum_{i=1}^{N_{S}} \left\lVert  \hat{\rho}_i - \rho_i \right\rVert_1, \quad \text{with} \quad \rho_i \in \text{Separable},
\end{equation}
where $\hat{\rho}_i = D_{\theta}(E(\rho_i))$ denotes the estimated density matrix, $N_S$ is the size of the data set, and  $\| \cdot \|_1$ denotes the mean element-wise absolute difference (L1 Loss). We adopted this loss function for its simplicity and computational efficiency, which proved highly effective in our experiments. Our training objective is to learn the mathematical structure of separable states such that the autoencoder achieves low reconstruction error exclusively for this class:
\begin{equation}
\label{eq:threshold1}
\left\lVert  \hat{\rho} - \rho \right\rVert_1 < \epsilon_d, \quad \forall , \rho \in \text{Separable},
\end{equation}
while simultaneously producing high reconstruction error for entangled states:
\begin{equation}
\label{eq:threshold2}
\left\lVert  \hat{\sigma} - \sigma \right\rVert_1 \geq \epsilon_d, \quad \forall , \sigma \in \text{Entangled}.
\end{equation}
The threshold $\epsilon_d$ is found during training, where $d$ is the local dimension of the bipartite system $d\times d$.  At the end of each epoch, we generate a set $\{ \rho_i \}_{i=1}^{N_{\epsilon}}$ of separable states and a set $\epsilon_d = \text{max}\{ \left\lVert  \hat{\rho}_i - \rho_i \right\rVert_1 \}_{i=1}^{N_{\epsilon}}$.

\subsection*{Data generation}

Our training dataset contains separable states only. As we study here bipartite systems, we generate the states for our training set according to
\begin{equation}
    \label{eq:bipartite_sep}
     \rho = \sum_{i=1}^{M} p_i \rho^A_i \otimes \rho^B_i.
\end{equation}
First, we choose a $M$-dimensional probability distribution $p = (p_1, \ldots, p_M)^T$ such as $\sum_{i=1}^{M}p_i = 1$, with $1\leq M \leq M_{\text{max}}$ for a predefined $M_{\text{max}}$. Then, each $\rho^A_i$ and $\rho^B_i$ in Eq.~(\ref{eq:bipartite_sep}) are randomly chosen according to the Hilbert-Schmidt measure. 

For \textit{quantum discord classification}, we employ the same autoencoder architecture but train it on classical-classical states Eq.~(\ref{eq:cc}) rather than separable states Eq.~(\ref{eq:bipartite_sep}). In these cases, the state $\rho_i \in \textit{States with discord}$ Eq.~(\ref{eq:norm}) while the threshold conditions become $\rho \in \textit{States without discord}$ and $\sigma \in \textit{States with discord}$. For generating classical-classical, quantum-classical, and classical-quantum states, unitary bases are randomly generated according to the Haar-Measure. Each column of these unitary transformations is then used as a basis vector to construct the corresponding quantum states.

\subsection*{Bound entangled state generation}
We can leverage our classification method to generate novel bound entangled density matrices. To achieve this, we model bipartite density matrices as a mixture of states represented by a neural network:
\begin{equation}
\label{eq:neural_network_state}
\rho_{\phi} = \sum_{j=1}^{\kappa} w_j \rho_{\vec{\phi}_j} 
\end{equation}
where $\phi = \{\vec{\phi}_k, w_k\}$ represents the network's weights. The coefficients $w_j$ are trainable parameters satisfying $w_j \geq 0$ and $\sum_j w_j = 1$, and $\vec{\phi}_j$ represents the set of trainable parameters for each $ \rho_{\vec{\phi}_j} $. Modeling the state as a neural network allows us to use deep learning frameworks powered by highly optimized linear algebra and optimization libraries. Thus, to generate novel instances of bound entangled states, we need to find the parameters $\phi$ that maximizes the following objective function:
\begin{equation}
    \label{eq:cost_bound_ent_generation}
   \max_{\phi}\Big( \lambda\,\big( \left\lVert \hat{\rho}_{\phi}-\rho_{\phi}\right\rVert_{1} - \epsilon_d \big) \;-\; \left\lVert \rho_{\phi} - \rho_{\phi}^{T_A} \right\rVert_{1} \Big)
\end{equation}
where
\begin{equation}
\lambda = \begin{cases} 1, & \text{if } \bigl\lVert \hat{\rho}_{\phi} - \rho_{\phi} \bigr\rVert_{1} < \epsilon_d,\\ 0, & \text{otherwise.} \end{cases}    
\end{equation}
The first term in Equation (\ref{eq:cost_bound_ent_generation}) drives the optimizer to find a quantum state whose reconstruction error is larger than the threshold $\epsilon_d$. The second term enforces the state to have a partial transpose close to the state itself, steering $\rho_{\phi}$ toward being a PPT state. As the first term can become too large, the penalty $\lambda$ helps the optimizer concentrate on minimizing the second term once the first term is large enough.

The quantum states $\rho_{\vec{\phi}_j}$, $j=1,\dots,\kappa$, in Equation (\ref{eq:cost_bound_ent_generation}) are valid by construction, i.e., $\rho_{\vec{\phi}_i} \geq 0$ and $\text{Tr}(\rho_{\vec{\phi}_i}=1$. 
For each branch $j \in \{1, \dots, M\}$, the network outputs a complex matrix $H_j \in \mathbb{C}^{d \times d}$:
\begin{equation}
H_j = \mathcal{R}_j(\mathbb{\mathbb{1}}) + i\, \mathcal{I}_j(\mathbb{1})
\end{equation}
where $i = \sqrt{-1}$, and $\mathcal{R}_j$ and $\mathcal{I}_j$ are the $j$-th linear transformations for the real and imaginary parts, respectively, acting on the identity operator $\mathbb{1}$. We then form a PSD matrix via the product:
\begin{equation}
\label{eq:gram_matrix}\tilde{\rho}_{\vec{\phi}_j} = H_j H_j^\dagger
\end{equation}
To satisfy the unit trace condition required for a density matrix, we normalize each term:
\begin{equation}
\rho_{\vec{\phi}_j} = \frac{ \tilde{\rho}_{\vec{\phi}_j} }{\text{Tr}(\tilde{\rho}_{\vec{\phi}_j})}
\end{equation}
Finally, the weights $w_j$ are determined by a Softmax layer, $w_j = \text{softmax}(z)_j$, where $z$ is the output of a linear layer. This ensures that the total state $\rho_{\phi}$ from Equation (\ref{eq:neural_network_state}) is a valid convex combination of density matrices.

\section{Results}   
In this section, we evaluate the performance of our classification and bound-entangled state generation method. For the classification task, we focus on accuracy (success rate) and the robustness of the classifiers under local unitary transformations. For each dimension $d \in \{3, \dots, 7\}$, we trained CAEs using the Adam optimizer with a learning rate of $10^{-4}$. Regularization hyperparameters, including dropout and L1 strength, were individually optimized based on validation performance. Training dataset sizes varied across dimensions, ranging from $50000$ to $200000$ samples with $M_{max} = 2$ (see Equation (\ref{eq:bipartite_sep})) for all cases. The model was implemented in PyTorch, and full training specifications are available on GitHub \cite{github_danuzco}.

\subsection*{Entanglement classification}

\begin{figure}[htbp]
    \centering
    \includegraphics[scale=0.445]{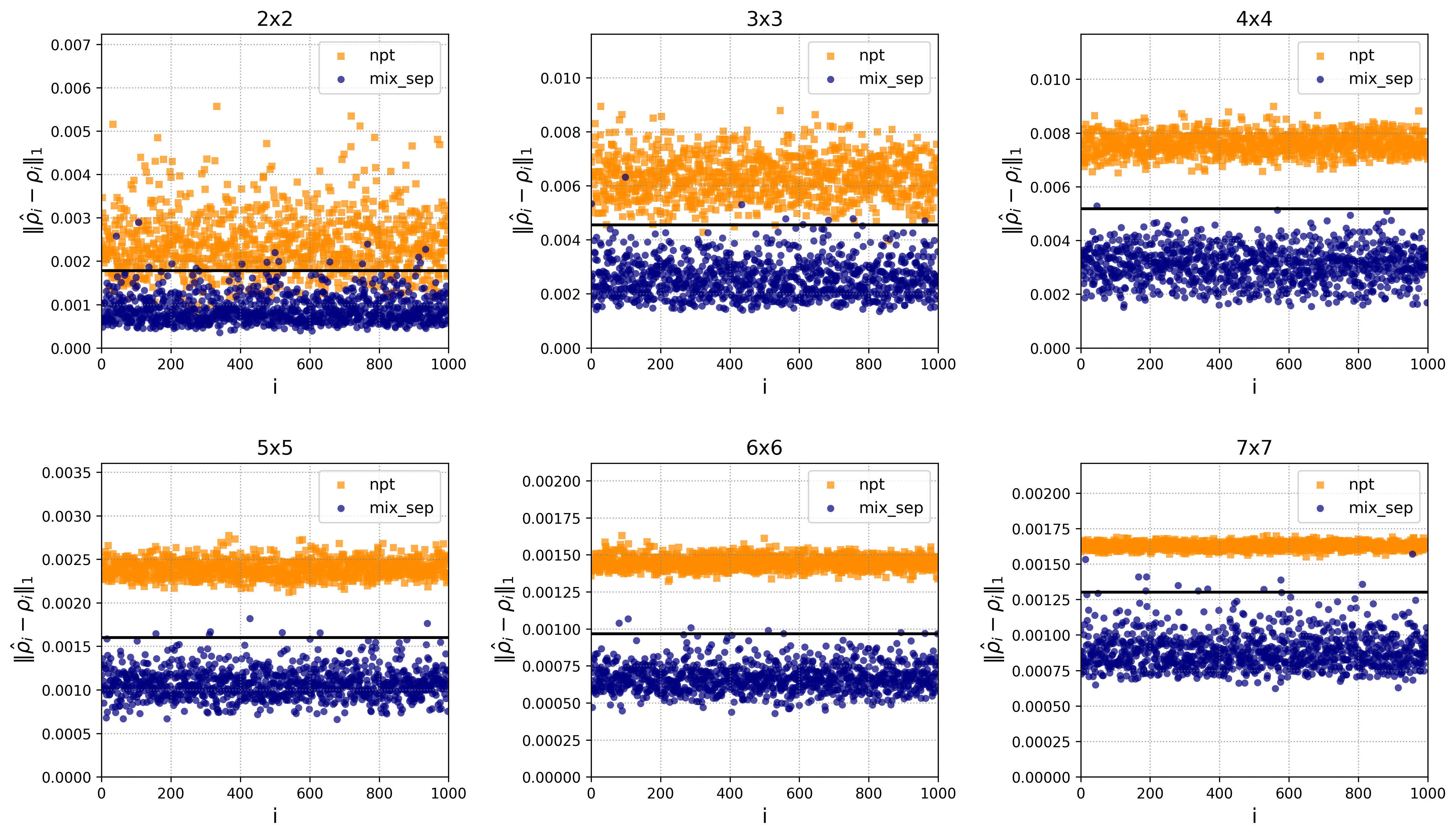}
    \caption{Reconstruction errors $\left\lVert  \hat{\rho}_i - \rho_i \right\rVert_1$ of bipartite $d\times d$ quantum states. Orange squares correspond to NPT states whereas blue circles correspond to separable states generated according to Equation (\ref{eq:bipartite_sep}).The horizontal black line indicates the decision threshold $\epsilon_d$; reconstruction errors below this line are classified as separable, while those above are identified as non-separable.  }
    \label{fig:entanglement_classification}
\end{figure}

For inference, we generated a validation data set $\{ \rho_i \}$ formed by $1000$ samples of NPT states and $1000$ samples of mixed separable ($\textit{mix\_sep}$) states with $M_{max} = 5$. These samples were reconstructed using the CAE and the L1 Loss was calculated for each reconstruction. Figure~\ref{fig:entanglement_classification} shows the reconstruction error $\|  \hat{\rho}_i - \rho_i \|_{1}$ for each sample $i$ across different $d\times d$ cases. In all cases, we observe a clear separation between NPT states (orange circles) and separable states (blue circles). The horizontal black line represents the empirically determined threshold $\epsilon_d$ for each dimension $d$: samples above this line are classified as entangled, while those below are classified as separable.

\subsection*{Entanglement classification under local unitary transformations}
\begin{figure}[htbp]
    \centering
    \includegraphics[scale=0.445]{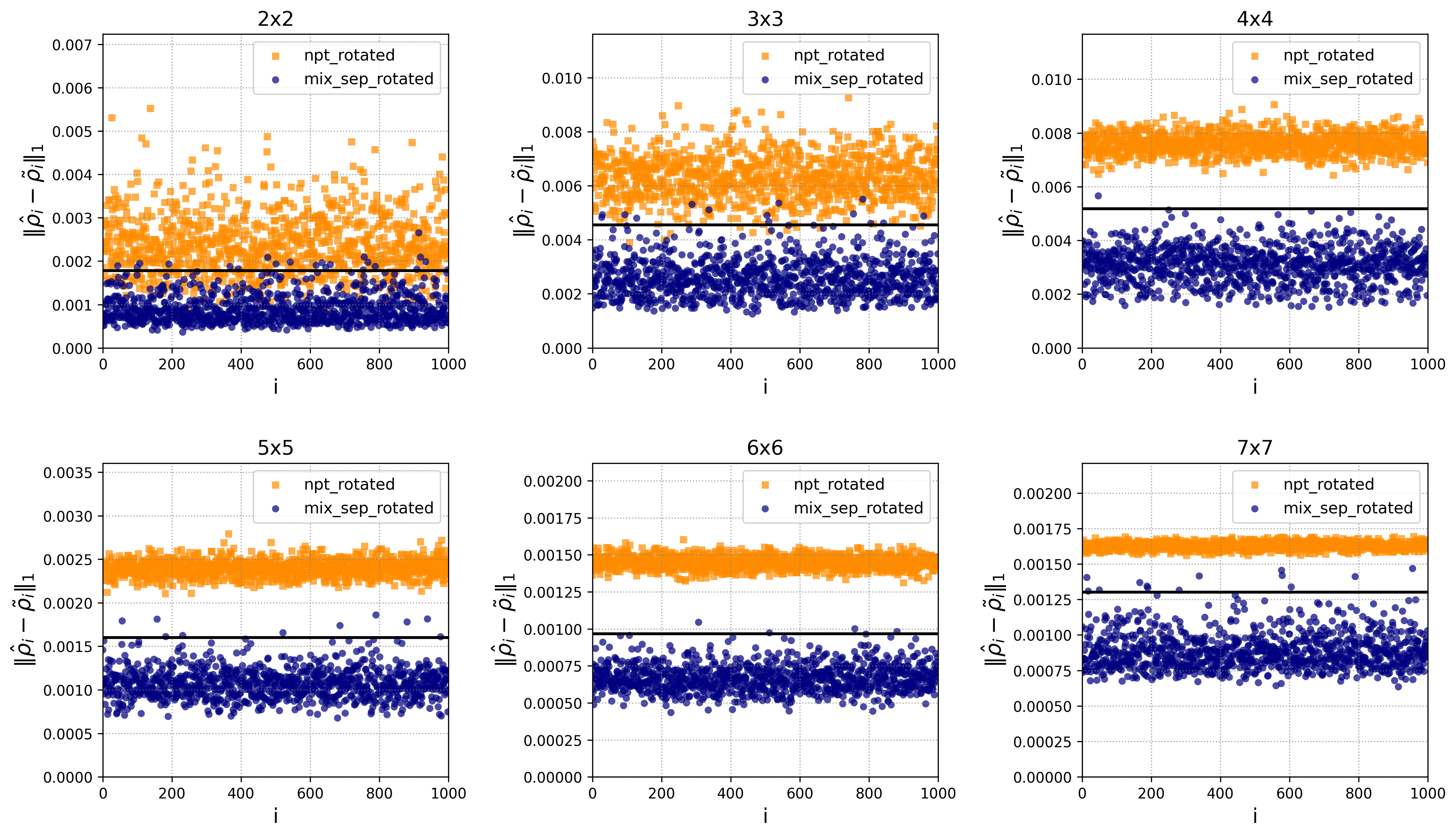}
    \caption{Reconstruction errors $\left\lVert  \hat{\rho}_i - \tilde{\rho}_i \right\rVert_1$ of the validation set after applying local unitaries $\tilde{\rho}_i =U_A \otimes U_B\left( \rho_i \right) U^{\dagger}_A \otimes U^{\dagger}_B$. Orange squares correspond to NPT states whereas blue circles correspond to separable states generated according to Equation (\ref{eq:bipartite_sep}).}
    \label{fig:entanglement_classification_rotated}
\end{figure}

\begin{figure}[htbp]
    \centering
    \includegraphics[scale=0.6]{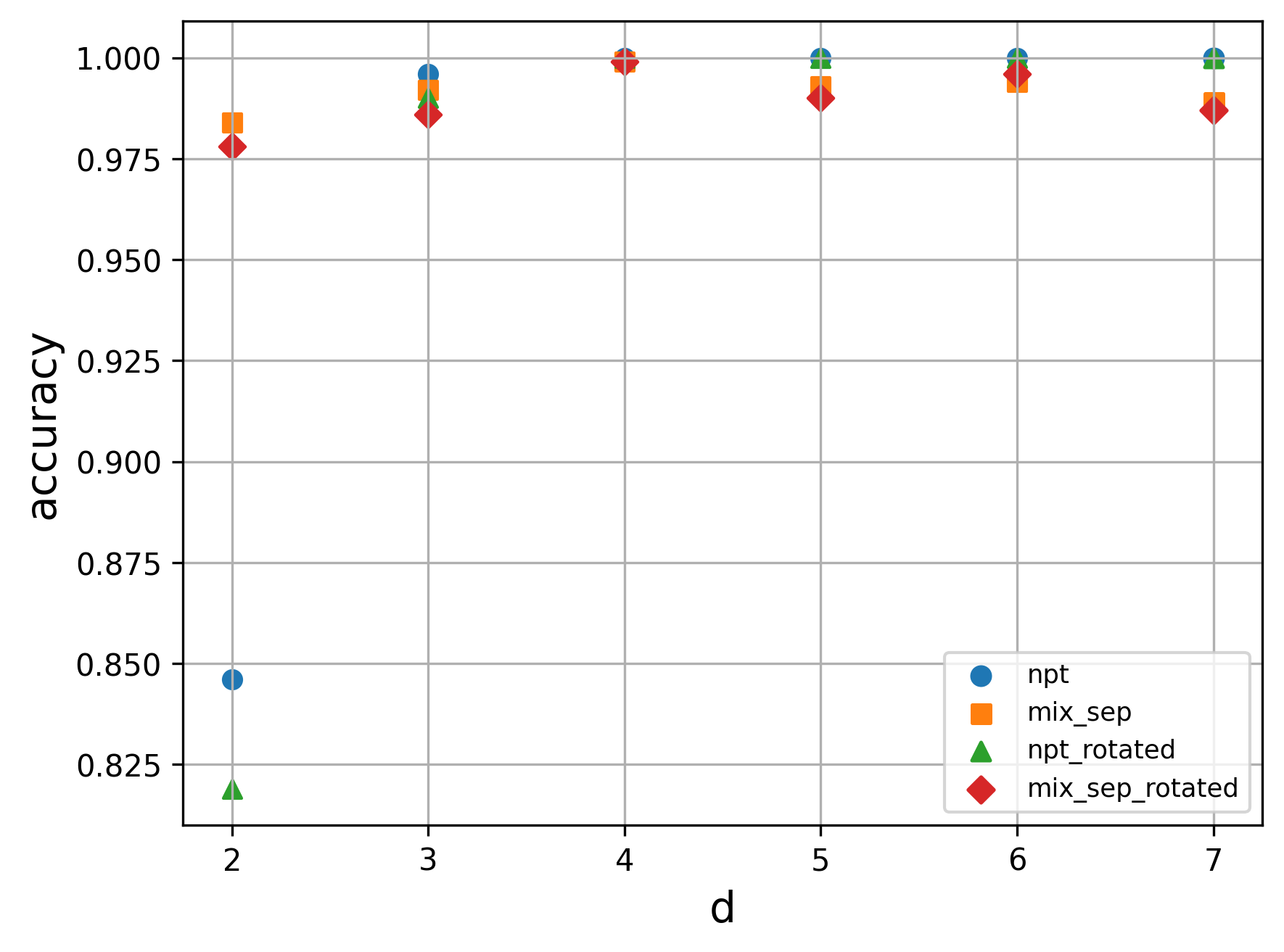}
    \caption{Validation accuracy for entanglement classification as a function of dimension $d$. The model demonstrates robust performance ($>98\%$) for most states, with the exception of NPT and NPT-rotated (after applying local unitaries) in the $d=2$ bipartite system. }
    \label{fig:accuracy}
\end{figure}

We also reconstructed the density matrices from the validation set after applying local unitary transformations to each state. Figure \ref{fig:entanglement_classification_rotated} displays $\left\lVert  \hat{\rho}_i - \tilde{\rho}_i \right\rVert_1$, where $\tilde{\rho}_i =U_A \otimes U_B\left( \rho_i \right) U^{\dagger}_A \otimes U^{\dagger}_B$ represents the rotated state and $\hat{\rho}_i = D_{\theta}( E( \tilde{\rho}_i ))$ is its reconstruction. The results closely resemble those in Figure \ref{fig:entanglement_classification}, demonstrating that the model maintains consistent performance under local unitary transformations for these state types. As shown in Figure \ref{fig:accuracy}, the classification accuracy remains nearly identical to that achieved with the original unrotated samples. The model exhibits accuracies above $98\%$ (except for NPT states in $2\times2$ systems). Applying local unitaries might be helpful for classifying uncommon quantum states, as states with bound entanglement. 

\subsection*{Entanglement classification for bound entangled states}
\begin{figure}
    \centering
    \includegraphics[scale=0.445]{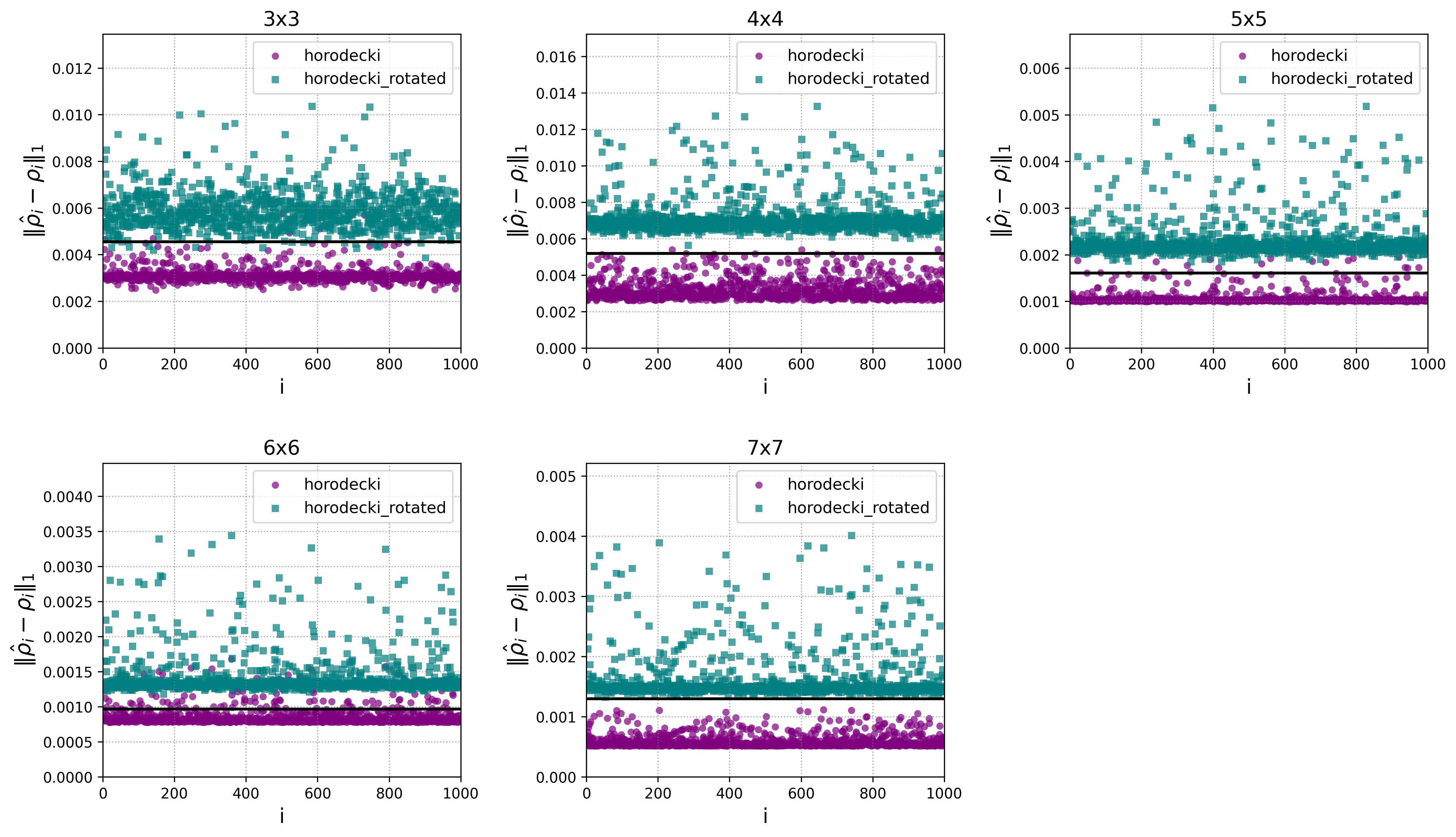}
    \caption{Reconstruction errors $\left\lVert  \hat{\rho}_i - \rho_i \right\rVert_1$ for samples of the Horodecki-like family of bound entangled states before and after local unitaries are applied.}
    \label{fig:horodecki_dxd}
\end{figure}

Figure \ref{fig:horodecki_dxd} shows the reconstruction error before and after applying local unitaries to the Horodecki-like family \cite{CHRUSCINSKI20112793} for $d \in \{3, \dots, 7\}$, which generalizes the two-qutrit Horodecki family of bound entangled states \cite{horodecki_1998_bound_entanglement}. We observe that these types of states are misclassified when reconstructed in their original basis. However, after applying local unitaries, the reconstruction errors significantly increase, moving into the entangled region and enabling proper classification. Since our autoencoders were trained exclusively on separable states, they successfully learned local basis invariance. However, because bound entangled states reside in a region of the state space geometrically adjacent to the separable set \cite{Bengtsson_Zyczkowski_2006}, local unitaries serve to expose the underlying entangled nature of the state.

\begin{figure}%
    \centering
    \subfloat[\centering]{{\includegraphics[scale=0.425]{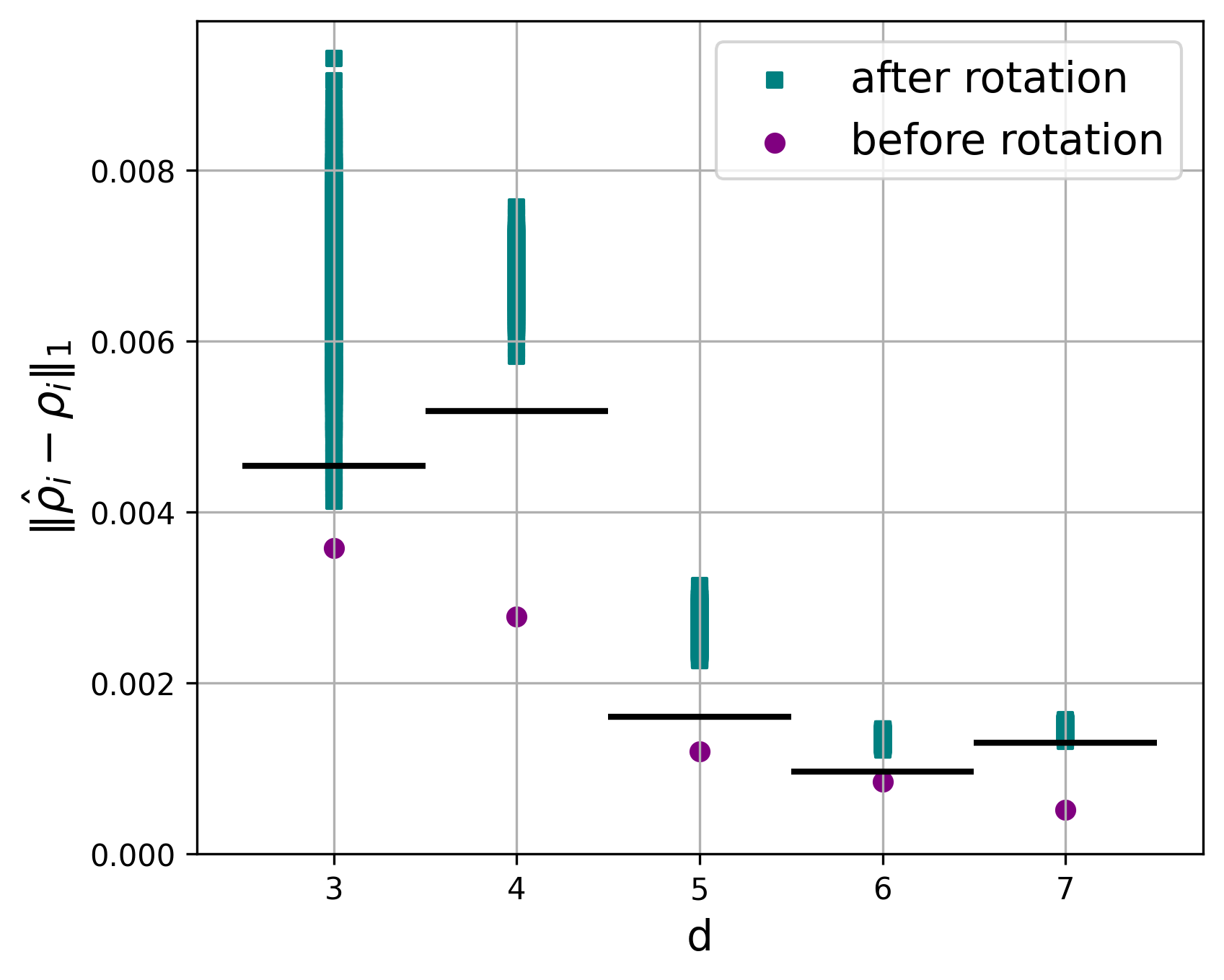} }}%
    \qquad
    \subfloat[\centering]{{\includegraphics[scale=0.425]{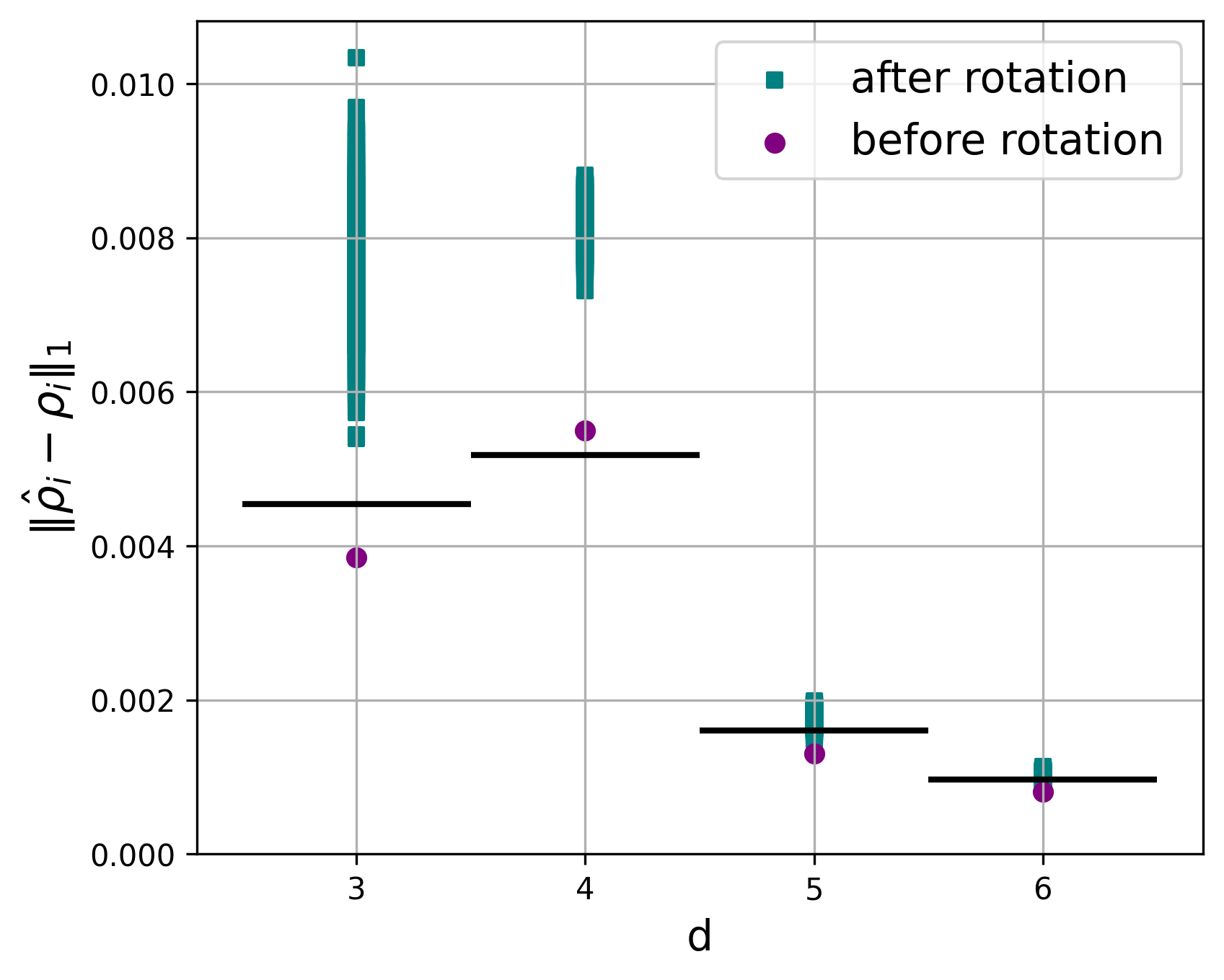} }}%
    \caption{$\left\lVert  \hat{\rho}_i - \rho_i \right\rVert_1$ vs $d$ before and after applying $1000$ local transformations to density matrices of a) the Horodecki-like family and b) UPB-based density matrices. Black horizontal lines correspond to $\epsilon_d$.}%
    \label{fig:repeated_unitaries}%
\end{figure}

As shown in Figure \ref{fig:horodecki_dxd}, our method achieves perfect accuracy in detecting entanglement for this family across all tested dimensions, with the exception of the $3 \times 3$ case, which remains highly accurate at approximately $98\%$. We observed a similar trend when analyzing bound entangled states constructed from Unextendible Product Bases (UPB) \cite{bennett_1999} for $d \in \{3, \dots, 6\}$; in these instances, the application of local unitaries again shifted the reconstruction errors above the threshold, enabling successful detection. This improvement in classification is consistent, as illustrated in Figure \ref{fig:repeated_unitaries}, which shows the reconstruction error before and after local unitaries are applied for one instance of a Horodecki-like bound entangled state for each $d$.

This improvement in classification in consistent, as illustrated in Figure \ref{fig:repeated_unitaries}, which shows, for each $d$, the reconstructions error before and after a thousand random local unitaries are applied to both instances of a Horodecki-like bound entangled state and density UPB-based density matrices.


Despite the robustness of our method, certain null-measure families remain challenging to detect, even when included in the training set. For instance, classical-classical, quantum-classical, and classical-quantum states are frequently misclassified and remain so even after the application of local unitaries. This behavior is expected, as these states constitute a vanishingly small volume of the separable state manifold \cite{Bengtsson_Zyczkowski_2006}. However, this does not represent a significant limitation in practice, as these states are computationally trivial to classify using standard analytical methods. Our approach is instead optimized for the more challenging task of detecting entanglement in states where such simple separability criteria fail.

\subsection*{Bound entangled state generation}
We successfully generated novel bound entangled density matrices for dimensions $d \in \{3, \dots, 7\}$. Specifically, we generated PPT density matrices with reconstruction errors that exceed the threshold $\epsilon_d$. To maximize Equation (\ref{eq:cost_bound_ent_generation}), we employed the Adam optimizer with a learning rate of $2 \times 10^{-4}$. We observed that the optimal number of mixture components scaled with dimension: while $\kappa=3$ was sufficient for $3 \leq d \leq 5$, convergence was improved by setting $\kappa=4$ and $\kappa=5$ for $d=6$ and $d=7$, respectively. We verified the entanglement of the generated states using the Symmetric Extension criteria \cite{doherty_2004}, implemented through the QETLAB library \cite{qetlab}.

\subsection*{Discord detection}
\begin{figure}
    \centering
    \includegraphics[scale=0.445]{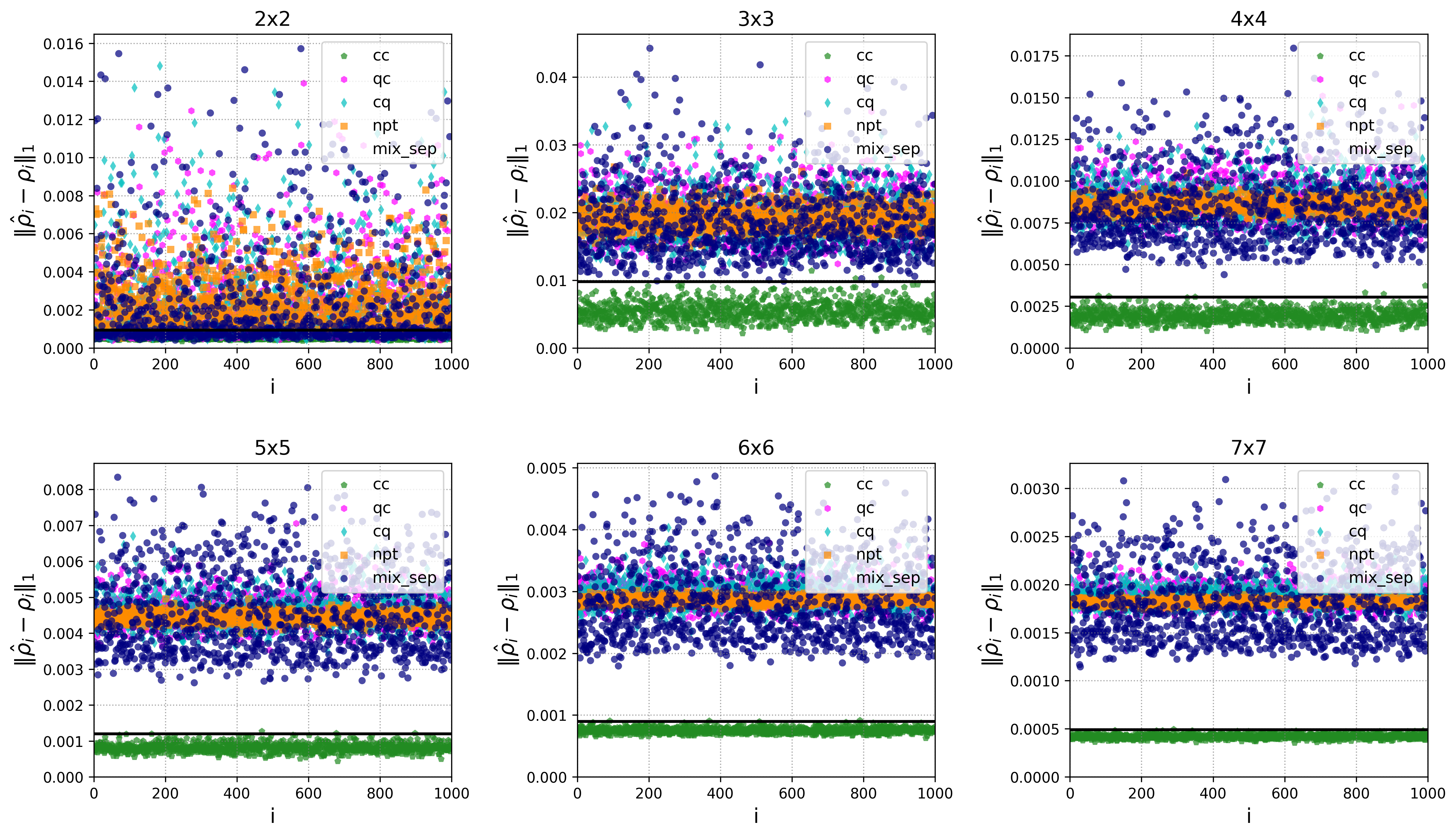}
    \caption{Reconstruction errors $\left\lVert  \hat{\rho}_i - \rho_i \right\rVert_1$ of bipartite $d\times d$ quantum states. Reconstruction errors below the horizontal black line are classified as states with zero discord, while those above are identified as states with discord. }
    \label{fig:discord_classification}
\end{figure}

\begin{figure}
    \centering
    \includegraphics[scale=0.6]{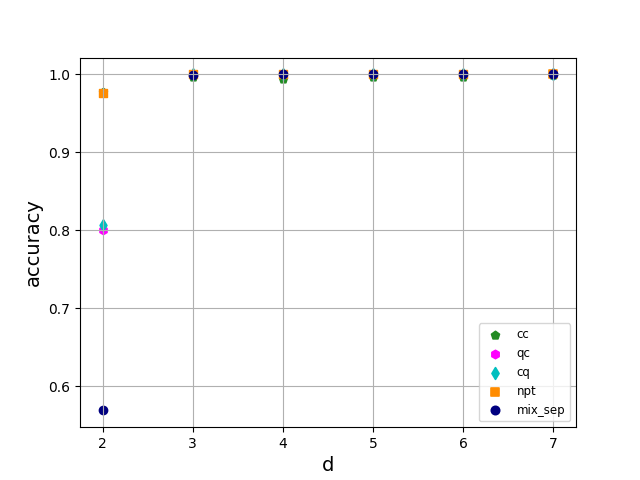}
    \caption{Validation accuracy for discord classification as a function of dimension $d$. The model exhibits a high performance, with perfect accuracies for all the cases except for $d=2$.}
    \label{fig:accuracy_discord}
\end{figure}

Training CAEs to distinguish states with discord from those without required significantly fewer computational resources than in the entanglement case. Specifically, the models achieved high performance with fewer trainable parameters and a smaller number of training samples. Figure \ref{fig:discord_classification} shows the reconstruction error for $1000$ samples of each type of state in dimensions $d \in \{3, \dots, 7\}$. The CAEs clearly separates classical-classical states (zero discord) from states possessing discord, such as quantum-classical, classical-quantum, and the mixed separable states generated via Equation (\ref{eq:bipartite_sep}). As shown in Figure \ref{fig:accuracy_discord}, the classification accuracy reaches nearly $100\%$ for all cases except $d=2$. 

\section{Conclusions and discussion}

In this work, we explored the use of CAEs for entanglement and discord classification in bipartite quantum systems. Our numerical studies show that CAEs can capture the mathematical structure of generic bipartite separable states, enabling thus to decide, with high probability, whether a given bipartite density matrix is separable or not. Our method achieves classification accuracies that exceed $98\%$ for $d\times d$ systems with $d \in \{3, \dots, 7\}$, successfully distinguishing separable states from both NPT and bound entangled states. Interestingly, our numerical simulations indicate that the CAEs generates a gap in the value of the loss function between separable and entangled states. This gap becomes more visible as the dimension $d$ increases.

The proposed method exhibits significant robustness under the action of local unitary operations, particularly in the classification of mixed separable and NPT states. Although this robustness does not inherently extend to specific families of bound entangled states, the application of local unitary transformations proved to be a constructive step in enhancing classification accuracy for these cases. Furthermore, our results indicate that local unitary transformations do not lead to a significant reduction in the reconstruction error for separable or entangled density matrices, ensuring that the classification of generic states remains uncompromised. Although certain null-measure separable states—such as quantum-classical and classical-quantum states—may be misclassified, this does not represent a fundamental limitation of the method, as discussed in the preceding results section.


A key contribution of our work is a novel technique for generating bound entangled states—among the most elusive and analytically challenging classes of entangled quantum states. Using our trained classifier within a gradient-based optimization framework, we successfully generated and verified bound entangled bipartite density matrices with local dimensions ranging from three to seven.

For quantum discord, our model demonstrated remarkable efficiency and simplicity, achieving accurate classification in a single training epoch. Discord classification exhibits a similar phenomenon as in the case of entanglement: the CAEs generate a gap between the values of the loss function for states with and without discord.

\section*{Acknowledgments}
This work was supported by the National Agency for Research and Development (ANID) through grants FONDECyT Regular No.\thinspace1230586 and No.\thinspace1231940, ANID Grants No.\thinspace$2023-3230427$ and No.\thinspace$2022-21221096$, Anillo Temático ATE250066, and ANID Millennium Science Initiative Program $\text{ICN17}\_\text{012}$. 

\appendix
\section*{Appendix A: Models architectures}

This appendix provides the detailed specifications for the convolutional autoencoder architectures employed across the $d \times d$ bipartite systems. For each local dimension $d \in \{2, \dots, 7\}$, we report the precise sequence of layers and the corresponding hyperparameters utilized in our numerical implementation.The models follow a consistent encoder–bottleneck–decoder topology. To ensure reproducibility, we explicitly list the parameters for batch normalization, activation functions (LeakyReLU and GELU), and spatial dropout in their exact order of execution within the computational graph.

\begin{table}[htb!]
\centering
\caption{Architecture for bipartite system with local dimension $d=2$.}
\label{tab:arch_d2}
\small
\begin{tabular}{ll}
\hline
\textbf{Layer} & \textbf{Configuration / Hyperparameters} \\
\hline
\textit{Encoder} & \\
Conv2D & $in=2, out=200, \text{Kernel}=2\times2, \text{Stride}=2\times2, \text{Pad}=0$ \\
BatchNorm2D & $channels=200, \epsilon=10^{-5}$ \\
LeakyReLU & $\text{Negative Slope}=0.01$ \\
Dropout2D & $\text{Dropout Rate}=0.5$ \\
\hline
Conv2D & $in=200, out=133, \text{Kernel}=2\times2, \text{Stride}=2\times2, \text{Pad}=0$ \\
BatchNorm2D & $channels=133$ \\
GELU & -- \\
Dropout2D & $\text{Dropout Rate}=0.5$ \\
\hline
\textbf{Latent BatchNorm2D} & $channels=133$ \\
\hline
\textit{Decoder} & \\
ConvTranspose2D & $in=133, out=200, \text{Kernel}=2\times2, \text{Stride}=2\times2, \text{Pad}=0$ \\
BatchNorm2D & $channels=200$ \\
GELU & -- \\
Dropout2D & $\text{Dropout Rate}=0.5$ \\
\hline
ConvTranspose2D & $in=200, out=2, \text{Kernel}=2\times2, \text{Stride}=2\times2, \text{Pad}=0$ \\
\hline
\end{tabular}

\vspace{0.8cm}

\caption{Architecture for bipartite system with local dimension $d=3$.}
\label{tab:arch_d3}
\begin{tabular}{ll}
\hline
\textbf{Layer} & \textbf{Configuration / Hyperparameters} \\
\hline
\textit{Encoder} & \\
Conv2D & $in=2, out=150, \text{Kernel}=3\times3, \text{Stride}=2\times2, \text{Pad}=1$ \\
BatchNorm2D & $channels=150, \epsilon=10^{-5}$ \\
LeakyReLU & $\text{Negative Slope}=0.01$ \\
Dropout2D & $\text{Dropout Rate}=0.2$ \\
\hline
Conv2D & $in=150, out=100, \text{Kernel}=3\times3, \text{Stride}=2\times2, \text{Pad}=1$ \\
BatchNorm2D & $channels=100$ \\
GELU & -- \\
Dropout2D & $\text{Dropout Rate}=0.2$ \\
\hline
Conv2D & $in=100, out=75, \text{Kernel}=3\times3, \text{Stride}=2\times2, \text{Pad}=1$ \\
BatchNorm2D & $channels=75$ \\
LeakyReLU & $\text{Negative Slope}=0.01$ \\
Dropout2D & $\text{Dropout Rate}=0.2$ \\
\hline
\textbf{Latent BatchNorm2D} & $channels=75$ \\
\hline
\textit{Decoder} & \\
ConvTranspose2D & $in=75, out=100, \text{Kernel}=3\times3, \text{Stride}=2\times2, \text{Pad}=1$ \\
BatchNorm2D & $channels=100$ \\
GELU & -- \\
Dropout2D & $\text{Dropout Rate}=0.2$ \\
\hline
ConvTranspose2D & $in=100, out=150, \text{Kernel}=3\times3, \text{Stride}=2\times2, \text{Pad}=1$ \\
BatchNorm2D & $channels=150$ \\
LeakyReLU & $\text{Negative Slope}=0.01$ \\
Dropout2D & $\text{Dropout Rate}=0.2$ \\
\hline
ConvTranspose2D & $in=150, out=2, \text{Kernel}=3\times3, \text{Stride}=2\times2, \text{Pad}=1$ \\
\hline
\end{tabular}
\end{table}

\newpage

\begin{table}[htb!]
\centering
\caption{Architecture for bipartite system with local dimension $d=4$.}
\label{tab:arch_d4}
\small
\begin{tabular}{ll}
\hline
\textbf{Layer} & \textbf{Configuration / Hyperparameters} \\
\hline
\textit{Encoder} & \\
Conv2D & $in=2, out=200, \text{Kernel}=4\times4, \text{Stride}=2\times2, \text{Pad}=1$ \\
BatchNorm2D & $channels=200, \epsilon=10^{-5}$ \\
LeakyReLU & $\text{Negative Slope}=0.1$ \\
Dropout2D & $\text{Dropout Rate}=0.01$ \\
\hline
Conv2D & $in=200, out=100, \text{Kernel}=4\times4, \text{Stride}=2\times2, \text{Pad}=1$ \\
BatchNorm2D & $channels=100$ \\
GELU & -- \\
Dropout2D & $\text{Dropout Rate}=0.01$ \\
\hline
Conv2D & $in=100, out=66, \text{Kernel}=4\times4, \text{Stride}=2\times2, \text{Pad}=1$ \\
BatchNorm2D & $channels=66$ \\
LeakyReLU & $\text{Negative Slope}=0.1$ \\
Dropout2D & $\text{Dropout Rate}=0.01$ \\
\hline
\textbf{Latent BatchNorm2D} & $channels=66$ \\
\hline
\textit{Decoder} & \\
ConvTranspose2D & $in=66, out=100, \text{Kernel}=4\times4, \text{Stride}=2\times2, \text{Pad}=1$ \\
BatchNorm2D & $channels=100$ \\
GELU & -- \\
Dropout2D & $\text{Dropout Rate}=0.01$ \\
\hline
ConvTranspose2D & $in=100, out=200, \text{Kernel}=4\times4, \text{Stride}=2\times2, \text{Pad}=1$ \\
BatchNorm2D & $channels=200$ \\
LeakyReLU & $\text{Negative Slope}=0.1$ \\
Dropout2D & $\text{Dropout Rate}=0.01$ \\
\hline
ConvTranspose2D & $in=200, out=2, \text{Kernel}=4\times4, \text{Stride}=2\times2, \text{Pad}=1$ \\
\hline
\end{tabular}

\vspace{0.8cm}

\caption{Architecture for bipartite system with local dimension $d=5$.}
\label{tab:arch_d5}
\begin{tabular}{ll}
\hline
\textbf{Layer} & \textbf{Configuration / Hyperparameters} \\
\hline
\textit{Encoder} & \\
Conv2D & $in=2, out=200, \text{Kernel}=5\times5, \text{Stride}=2\times2, \text{Pad}=2$ \\
BatchNorm2D & $channels=200, \epsilon=10^{-5}$ \\
LeakyReLU & $\text{Negative Slope}=0.1$ \\
Dropout2D & $\text{Dropout Rate}=0.01$ \\
\hline
Conv2D & $in=200, out=100, \text{Kernel}=5\times5, \text{Stride}=2\times2, \text{Pad}=2$ \\
BatchNorm2D & $channels=100$ \\
GELU & -- \\
Dropout2D & $\text{Dropout Rate}=0.01$ \\
\hline
Conv2D & $in=100, out=66, \text{Kernel}=5\times5, \text{Stride}=2\times2, \text{Pad}=2$ \\
BatchNorm2D & $channels=66$ \\
LeakyReLU & $\text{Negative Slope}=0.1$ \\
Dropout2D & $\text{Dropout Rate}=0.01$ \\
\hline
\textbf{Latent BatchNorm2D} & $channels=66$ \\
\hline
\textit{Decoder} & \\
ConvTranspose2D & $in=66, out=100, \text{Kernel}=5\times5, \text{Stride}=2\times2, \text{Pad}=2$ \\
BatchNorm2D & $channels=100$ \\
GELU & -- \\
Dropout2D & $\text{Dropout Rate}=0.01$ \\
\hline
ConvTranspose2D & $in=100, out=200, \text{Kernel}=5\times5, \text{Stride}=2\times2, \text{Pad}=2$ \\
BatchNorm2D & $channels=200$ \\
LeakyReLU & $\text{Negative Slope}=0.1$ \\
Dropout2D & $\text{Dropout Rate}=0.01$ \\
\hline
ConvTranspose2D & $in=200, out=2, \text{Kernel}=5\times5, \text{Stride}=2\times2, \text{Pad}=2$ \\
\hline
\end{tabular}
\end{table}

\newpage

\begin{table}[htb!]
\centering
\caption{Architecture for bipartite system with local dimension $d=6$.}
\label{tab:arch_d6}
\small
\begin{tabular}{ll}
\hline
\textbf{Layer} & \textbf{Configuration / Hyperparameters} \\
\hline
\textit{Encoder} & \\
Conv2D & $in=2, out=80, \text{Kernel}=12\times12, \text{Stride}=1\times1, \text{Pad}=5$ \\
BatchNorm2D & $channels=80, \epsilon=10^{-5}$ \\
LeakyReLU & $\text{Negative Slope}=0.01$ \\
Dropout2D & $\text{Dropout Rate}=0.2$ \\
\hline
Conv2D & $in=80, out=40, \text{Kernel}=12\times12, \text{Stride}=3\times3, \text{Pad}=5$ \\
BatchNorm2D & $channels=40$ \\
GELU & -- \\
Dropout2D & $\text{Dropout Rate}=0.2$ \\
\hline
Conv2D & $in=40, out=26, \text{Kernel}=12\times12, \text{Stride}=1\times1, \text{Pad}=5$ \\
BatchNorm2D & $channels=26$ \\
LeakyReLU & $\text{Negative Slope}=0.01$ \\
Dropout2D & $\text{Dropout Rate}=0.2$ \\
\hline
\textbf{Latent BatchNorm2D} & $channels=26$ \\
\hline
\textit{Decoder} & \\
ConvTranspose2D & $in=26, out=40, \text{Kernel}=12\times12, \text{Stride}=1\times1, \text{Pad}=5$ \\
BatchNorm2D & $channels=40$ \\
GELU & -- \\
Dropout2D & $\text{Dropout Rate}=0.2$ \\
\hline
ConvTranspose2D & $in=40, out=80, \text{Kernel}=12\times12, \text{Stride}=3\times3, \text{Pad}=5$ \\
BatchNorm2D & $channels=80$ \\
LeakyReLU & $\text{Negative Slope}=0.01$ \\
Dropout2D & $\text{Dropout Rate}=0.2$ \\
\hline
ConvTranspose2D & $in=80, out=2, \text{Kernel}=12\times12, \text{Stride}=1\times1, \text{Pad}=5$ \\
\hline
\end{tabular}

\vspace{0.8cm}

\caption{Architecture for bipartite system with local dimension $d=7$.}
\label{tab:arch_d7}
\begin{tabular}{ll}
\hline
\textbf{Layer} & \textbf{Configuration / Hyperparameters} \\
\hline
\textit{Encoder} & \\
Conv2D & $in=2, out=70, \text{Kernel}=15\times15, \text{Stride}=2\times2, \text{Pad}=7$ \\
BatchNorm2D & $channels=70, \epsilon=10^{-5}$ \\
LeakyReLU & $\text{Negative Slope}=0.1$ \\
Dropout2D & $\text{Dropout Rate}=0.1$ \\
\hline
Conv2D & $in=70, out=35, \text{Kernel}=15\times15, \text{Stride}=4\times4, \text{Pad}=7$ \\
BatchNorm2D & $channels=35$ \\
GELU & -- \\
Dropout2D & $\text{Dropout Rate}=0.1$ \\
\hline
Conv2D & $in=35, out=23, \text{Kernel}=15\times15, \text{Stride}=2\times2, \text{Pad}=7$ \\
BatchNorm2D & $channels=23$ \\
LeakyReLU & $\text{Negative Slope}=0.1$ \\
Dropout2D & $\text{Dropout Rate}=0.1$ \\
\hline
\textbf{Latent BatchNorm2D} & $channels=23$ \\
\hline
\textit{Decoder} & \\
ConvTranspose2D & $in=23, out=35, \text{Kernel}=15\times15, \text{Stride}=2\times2, \text{Pad}=7$ \\
BatchNorm2D & $channels=35$ \\
GELU & -- \\
Dropout2D & $\text{Dropout Rate}=0.1$ \\
\hline
ConvTranspose2D & $in=35, out=70, \text{Kernel}=15\times15, \text{Stride}=5\times5, \text{Pad}=7$ \\
BatchNorm2D & $channels=70$ \\
LeakyReLU & $\text{Negative Slope}=0.1$ \\
Dropout2D & $\text{Dropout Rate}=0.1$ \\
\hline
ConvTranspose2D & $in=70, out=2, \text{Kernel}=15\times15, \text{Stride}=2\times2, \text{Pad}=13$ \\
\hline
\end{tabular}
\end{table}

\clearpage
\printbibliography
\end{document}